\documentclass[10pt]{article}
\usepackage[latin1]{inputenc}
\usepackage[english]{babel}

\usepackage{latexsym}
\usepackage{amsmath} 
\usepackage{amssymb}
\usepackage{theorem}

\newtheorem{prp}{Proposition}
\newtheorem{thm}{Theorem}
\newtheorem{crl}{Corollary}
\newtheorem{lm}{Lemma}

\newtheorem{df}{Definition}

{ \theorembodyfont{\rmfamily}
\newtheorem{obs}{Remark}
\newtheorem{ej}{Example}
}

\newenvironment{dm}
  {\begin{trivlist}
    \item[\textit{\quad\quad\textsc{ Proof.}}]}
{$\Box$\end{trivlist}}

\newcommand{\C}{\mathbf C}

\begin{document}

\title{Non-integrability of some Hamiltonians with rational potentials}
\author{Primitivo Acosta-Humánez \& David Blázquez-Sanz}

\maketitle

\begin{abstract}
In this work we compute the families of classical Hamiltonians in
two degrees of freedom in which the Normal Variational Equation
around an invariant plane falls in Schr\"odinger type with
polynomial or trigonometrical potential. We analyze the
integrability of Normal Variational Equation in Liouvillian sense
using the Kovacic's algorithm. We compute all Galois groups of
Schr\"odinger type equations with polynomial potential.
 We also introduce a method of algebrization that transforms equations with transcendental
coefficients in equations with rational coefficients without
changing essentially the Galoisian structure of the equation. We
obtain Galoisian obstructions to existence of a rational first
integral of the original Hamiltonian via Morales-Ramis theory.
\end{abstract}

\begin{center}
{\bf Key Words:} Picard-Vessiot Theory, Hamiltonian systems, Integrability, Kovacic's Algorithm.
\end{center}


\tableofcontents

\section{Introduction}

  In \cite{MorSimo}, C. Sim\'o and J. Morales-Ruiz find the
complete list of two degrees of freedom classical hamiltonians, with
an invariant plane $\Gamma=\{x_2=y_2=0\}$ with Normal Variational
Equation of Lam\'e type along generic curves in $\Gamma$. In those
computations they use systematically differential equations
satisfied by coefficients of Lam\'e equation. It allow us to develop
a method to find the families of hamiltonians with invariant plane
$\Gamma$, with the property of having NVE satisfying certain
conditions. It allow us to find the families of hamiltonians with
generic NVE of type Mathieu, Shr\"odinger with polynomial potential
of odd degree, quantum harmonic oscillator, and some other simpler
examples.

  Then, we analyze those families of linear differential equations
in the context of Picard-Vessiot theory. For equations with rational
coefficients, we use Kovacic's Algorithm. In particular we give a
complete descripion of Galois groups of Schr\"odinger type equations
with polynomial potential (theorem \ref{polynint}). For equations
with transcendental coefficients, we develop a method of
algebrization. We characterize equations that can be algebrized
through change of variables of certain type (hamiltonian). We prove
the following result.

\medskip

\emph{
{\bf Algebrization algorithm.}
The differential equation
$\ddot{y}=r(t)y$
is algebrizable through a hamiltonian change of variable $x=x(t)$
if and only if there exists $f,\alpha$ such that
${\alpha'\over\alpha},\quad {f\over \alpha}\in \mathbf{C}(x),\text{ where } f(x(t))=r(t),
\quad \alpha(x)=2(H-V(x))=\dot x^2.$
Furthermore, the algebraic form of the equation $\ddot{y}=r(t)y$
is
\begin{equation*}
y''+{1\over2}{\alpha'\over \alpha}y'-{f\over\alpha}y=0.
\end{equation*}}

\medskip

  Once we get the complete analysis of linearized equations, we apply a theorem of
Morales-Ramis, obtaining the following results on the non integrability of those hamiltonians
for generic values of the parameters.

\medskip

{\bf Non-integrability Results.} \emph{ Hamiltonians,}
\begin{enumerate}
\item $\frac{y_1^2+y_2^2}{2} +
\frac{\lambda_4}{(\lambda_2 + 2\lambda_3 x_1)^2} + \lambda_0 -
\lambda_1 x_2^2 - \lambda_2 x_1x_2^2 - \lambda_3 x_1^2 x_2^2 +
\beta(x_1,x_2)x_2^3,$
with $\lambda_3\neq 0$;

\item $\frac{y_1^2+y_2^2}{2} + \lambda_0 + Q(x_1)x_2^2 +
\beta(x_1,x_2)x_2^3,$
where $Q(x_1)$ is a non-constant polynomial;

\item $\frac{y_1^2+y_2^2}{2}+\mu_0 + \mu_1x_1 +\frac{\omega^2
x_1^2}{2}- \lambda_0x_2^2 - \lambda_1x_1x_2^2 + \beta(x_1,x_2)x_2^3,$
with $\omega \neq 0$.

\item $\frac{y_1^2+y_2^2}{2} + \mu_0 +
\frac{\mu_1}{(\lambda_1+2\lambda_2x_1)^2}+\frac{\lambda_1\omega^2 x_1}{8\lambda_2}
+ \frac{\omega^2 x_1^2}{8} +  -\lambda_0 x_2^2 -  \lambda_1x_1x_2^2 - \lambda_2x_1^2x_2^2 +
\beta(x_1,x_2)x_2^3,$
with $\omega\neq 0$, and $\lambda_2\neq 0$.
\end{enumerate}
\emph{where $\beta(x_1,x_2)$ is any analytical function around $\Gamma = \{x_2 = y_2 = 0\}$,
do no admit any additional rational first integral.}

\medskip

{\bf Corollary.} \emph{Every integrable (by rational functions)
polynomial potentials with invariant plane $\Gamma=\{x_2=y_2=0\}$
can be written in the following form}

$$V=Q_1(x_1,x_2)x_2^3+\lambda_1x_2^2+\lambda_0,\quad \lambda_0,\lambda_1\in \mathbf{C}.$$

\medskip

Note that for $\beta(x_1, x_2)$ polynomial and some values of the
parameters we fall in the case of homogeneous polynomial
potentials. Those cases had been analyzed deeply in \cite{Mac,
Nak}.

\subsection{Picard-Vessiot theory}

The Picard-Vessiot theory is the Galois theory of linear
differential equations. In the classical Galois theory, the main
object is a group of permutations of the roots, while in the
Picard-Vessiot theory is a linear algebraic group. For polynomial
equations we want a solution in terms of radicals. From classical
Galois theory it is well known that this is possible if and only
if the Galois group is solvable.

An analogous situation holds for linear homogeneous differential
equations. For more details see \cite{VanSinger}. The following
definition is true for matrices $n\times n$, but for simplicity we
are restricting to matrices $2\times 2.$

\begin{df} An
algebraic group of matrices $2\times 2$ is a subgroup $G\subset
GL(2,\mathbf{C})$, defined by algebraic equations in its matrix
elements. That is, there exists a set of polynomials
$$\{P_i(x_{11},x_{12},x_{21},x_{22})\}_{i\in I},$$
such that
$$\left(\begin{array}{cc} x_{11} & x_{12} \\ x_{21} & x_{22}
\end{array}\right)\in G \quad\Leftrightarrow\quad \forall i\in I,
P_i(x_{11},x_{12},x_{21},x_{22}) = 0.$$
\end{df}
In this case we say that $G$ is an algebraic manifold provided of
an structure of group. In the remainder of this paper we only
work, as particular case, with linear differential equations of
second order
$$y''+ay'+by=0,\quad a,b\in \mathbf{C}(x).$$

Suppose that $y_1, y_2$ is a fundamental system of solutions of
the differential equation. This means that $y_1, y_2$ are linearly
independent over $\mathbf{C}$ and every solution is a linear
combination of these two. Let $L = \mathbf{C}(x)\langle y_1, y_2
\rangle = \mathbf{C}(x)(y_1, y_2, y_1', y_2')$, that is the
smallest differential field containing to $\mathbf{C}(x)$ and
$\{y_{1},y_{2}\}.$

\begin{df}[Differential Galois Group]
The group of all differential automorphisms of $L$ over
$\mathbf{C}(x)$ is called the {\emph Galois group} of $L$ over
$\mathbf{C}(x)$ and denoted by $Gal(L/\mathbf{C}(x))$ or also by
$Gal^L_{\mathbf{C}(x)}$. This means that for $\sigma\colon L\to
L$, $\sigma(a')=\sigma'(a)$ and  $\forall a\in \mathbf{C}(x),$
$\sigma(a)=a$.
\end{df}

If $\sigma \in Gal(L/\mathbf{C}(x))$ then $\sigma y_1, \sigma y_2$
is another fundamental system of solutions of the linear
differential equation. Hence there exists a matrix

$$A=
\begin{pmatrix}
a & b\\
c & d
\end{pmatrix}
\in GL(2,\mathbf{C}),$$ such that
$$\sigma
\begin{pmatrix}
y_{1}\\
y_{2}
\end{pmatrix}
=
\begin{pmatrix}
\sigma y_{1}\\
\sigma y_{2}
\end{pmatrix}
=A
\begin{pmatrix}
y_{1}\\
y_{2}
\end{pmatrix}
.$$ This defines a faithful representation
$Gal(L/\mathbf{C}(x))\to GL(2,\mathbf{C})$ and it is possible to
consider $Gal(L/\mathbf{C}(x))$ as a subgroup of
$GL(2,\mathbf{C})$. It depends on the choice of fundamental system
$y_1$, $y_2$, but only up to conjugate.

One of the fundamental results of the Picard-Vessiot theory is the
following theorem.

\begin{thm}  The Galois group $G=Gal(L/\mathbf{C}(x))$ is an
algebraic subgroup of $GL(2,\mathbf{C})$.
\end{thm}

Now we are interested in the differential equation
\begin{equation}\label{LDE}
\xi''=r\xi,\quad r\in \mathbf{C}(x).
\end{equation}

We recall that equation \eqref{LDE} can be obtained from the
general second order linear differential equation
$$y''+ay'+by=0,\quad a,b\in\mathbf{C}(x),$$
through the change of variable
$$y=e^{-{1\over 2}\int_{}^{}a}\xi,\quad r={a^2\over 4}+{a'\over 2}-b.$$

On the other hand, through the change of variable $v=\xi'/\xi$ we
get the associated Riccatti equation to equation \eqref{LDE}
\begin{equation}\label{Riccatti}
v'=r-v^2,\quad v={\xi'\over \xi}.
\end{equation}

For the differential equation \eqref{LDE},
$G=Gal(G/\mathbf{C}(x))$ is an algebraic subgroup
$SL(2,\mathbf{C})$.

Recall that an algebraic group $G$ has a unique connected normal
algebraic subgroup $G^0$ of finite index. This means that the
identity component $G^0$ is the biggest connected algebraic
subgroup of $G$ containing the identity.

\begin{df} Let $F$ be a differential extension of $\mathbf{C}(x)$, and let be $\eta$ solution of the differential equation
$$y''+ay'+by=0,\quad a,b\in F$$
\begin{enumerate}
\item $\eta$ is {\emph{algebraic}} over $F$ if $\eta$ satisfies a
polynomial equation with coefficients in $F$, i.e. $\eta$ is an
algebraic function of one variable.
\item $\eta$ is {\emph{primitive}} over $F$ if $\eta' \in F$,
i.e. $\eta = \int f$ for some $f \in F$.
\item $\eta$ is {\emph{exponential}} over $F$ if $\eta' /\eta \in F$,
i.e. $\eta = e^{\int f}$ for some $f \in F$.
\end{enumerate}
\end{df}

\begin{df}  A solution $\eta$ of the previous differential equation
is said to be {\emph{Liouvillian}} over $F$ if there is a tower of
differential fields
$$
    F = F_0 \subset F_1 \subset ... \subset F_m = L,
$$
with $\eta \in L$ and for each $i = 1,...,m$, $F_i =
F_{i-1}(\eta_i)$ with $\eta_i$ either algebraic, primitive, or
exponential over $F_{i-1}$. In this case we say that the
differential equation is integrable.
\end{df}

Thus a Liouvillian solution is built up using algebraic functions,
integrals and exponentials. In the case $F=\mathbf{C}(x)$ we get,
for instance logarithmic, trigonometric functions, but not special
functions such that the Airy functions.

We recall that a group $G$ is called solvable if and only if there
exists a chain of normal subgroups
$$e=G_0\triangleleft G_1 \triangleleft \ldots \triangleleft G_n=G$$ such that the
quotient $G_i/G_j$ is abelian for all $n\geq i\geq j\geq 0$.

\begin{thm}\label{LK}
The equation \eqref{LDE} is integrable (has Liouvillian solutions)
if and only if for $G=Gal(L/\mathbf{C}(x))$, the identity
component $G^0$ is solvable.
\end{thm}

Using the theorem \ref{LK}, Kovacic in 1986 introduced an
algorithm to solve the differential \eqref{LDE} and show that
\eqref{LDE} is integrable if and only if the solution of the
equation \eqref{Riccatti} is a rational function (case 1), is a
root of polynomial of degree two (case 2) or is a root of
polynomial of degree 4, 6, or 12 (case 3) (see Appendix A). Based
in the Kovacic's algorithm we have the following \emph{key}
result.

\begin{center}\emph{The Galois group of differential equation \eqref{LDE}
with $r=Q_k(x)$ a polynomial of degree $k>0$ is a non-abelian
connected group}.
\end{center}

For complete result, details and proof see Appendix A.

\subsection{Morales-Ramis theory}

  Morales-Ramis theory \cite{MorRamis1, MorRamis2}, see also \cite{MorMonograph}, relates the integrability of hamiltonian
systems to the integrability of linear differential equations. In
this approach analyze the linearization  (variational equations)
of hamiltonian systems along some known particular solution. If
the hamiltonian system is integrable, then we expect that the
linearized equation has good properties in the sense of
Picard-Vessiot theory. Exactly, for integrable hamiltonian
systems, the Galois group of the linearized equation must be
virtually abelian. It gives us the best non-integrability
criterion known for hamiltonian systems. This approach has been
extended to higher order variational equations in
\cite{MorRamSimo}.

\subsubsection{Integrability of hamiltonian systems}

  A symplectic manifold (real or complex), $M_{2n}$ is a $2n$-dimensional manifold,
provided with a closed 2-form $\omega$. This closed $2$-form gives
us a natural isomorphism  between vector bundles, $\flat\colon TM\to
T^*M$. Given a function $H$ on $M$, there is an unique vector field
$X_H$ such that,
  $$\flat(X_H) = dH$$
this is \emph{the hamiltonian vector field of $H$}. Furthermore, it
gives an structure of \emph{Poisson algebra} in the ring of
differentiable functions of $M_{2n}$ by defining:
$$\{H,F\} = X_H F.$$
  We say that $H$ and $F$ are \emph{in involution} if $\{H,F\} = 0$.
From our definition, it is obvious that $F$ is a \emph{first
integral} of $X_H$ if and only if $H$ and $F$ are in involution. In
particular $H$ is always a first integral of $X_H$. Moreover, if $H$
and $F$ are in involution, then their flows commute.

  The equations of the flow of $X_H$, in a system of canonical coordinates, $p_1,\ldots,p_n,q_1,\ldots,q_n$
(i.e. such that $\omega_2 = \sum_{i=1}^n p_i\wedge q_i$), are
written
$$\dot q = \frac{\partial H}{\partial p} \left( = \{H, q\}\right),
\quad \dot p  = - \frac{\partial H}{\partial q} \left(= \{ H,
p\}\right),$$ and they are known as \emph{Hamilton equations}.

\begin{thm}[Liouville-Arnold]
  Let $X_H$ be a hamiltonian defined on a real symplectic manifold
$M_{2n}$. Assume that there are $n$ functionally independent first
integrals $F_1,\ldots, F_n$ in involution.
  Let $M_a$ be a non-singular (i.e. $dF_1,\ldots, dF_n$ are independent  on very point of
$M_a$) level manifold, $$M_a = \{p\colon F_1(p)=a_1,\ldots,
F_n(p)=a_n\}.$$ Then,
\begin{enumerate}
\item If $M_a$ is compact and connected, then it is a torus
$M_a\simeq \mathbf R^n/\mathbf Z^n$.
\item In a neighborhood of the torus $M_a$ there are functions
$I_1,\ldots I_n,\phi_1,\ldots,\phi_n$ such that
$$\omega_2 = \sum_{i=1}^n d I_i\wedge d\phi_i,$$
and $\{H,I_j\} = 0$ for $j = 1,\ldots, n$.
\end{enumerate}
\end{thm}

  From now on, we will consider $\C^{2n}$ as a complex
symplectic manifold. Lioville-Arnold theorem gives us a notion of
integrability for Hamiltonian systems. A hamiltonian $H$ in
$\C^{2n}$ is called \emph{integrable in the sense of Liouville} if
there exist $n$ independent first integrals of $X_H$ in involution.
We will say that $H$ in integrable \emph{by terms of rational
functions} if  we can find those first integrals between rational
functions, or whatever.

\subsubsection{Variational equations}

We want to relate integrability of hamiltonian systems with
Picard-Vessiot theory. We deal with non-linear hamiltonian
systems. But, given a hamiltonian $H$ in $\C^{2n}$, and $\Gamma$
an integral curve of $X_H$, we can consider the \emph{first
variational equation} (VE), as
$$\mathcal{L}_{X_H}\xi=0,$$
in which the linear equation induced in the tangent bundle ($\xi$
represents a vector field supported on $\Gamma$).

 Let $\Gamma$ be parameterized by $\gamma\colon t\mapsto (x(t),y(t))$
in such way that
$$\frac{d x_i}{d t} = \frac{\partial H}{\partial y_i}, \quad
\frac{d y_i}{dt} = - \frac{\partial H}{\partial x_i}.$$

  Then the VE along $\Gamma$ is the linear system,
$$\left(\begin{array}{c} \dot \xi_i \\ \dot\eta_i, \end{array}\right) =
\left(\begin{array}{cc}
\frac{\partial^2 H}{\partial y_i\partial x_j}(\gamma(t)) & \frac{\partial^2 H}{\partial y_i \partial y_j}(\gamma(t)) \\
- \frac{\partial^2 H}{\partial x_i \partial x_j}(\gamma(t)) &
-\frac{\partial^2 H}{\partial x_i \partial y_j}(\gamma(t))
\end{array}\right)
 \left(\begin{array}{c}\xi_i
\\ \eta_i \end{array}\right).$$

  From the definition of Lie derivative, comes that
$$\xi_i(t) = \frac{\partial H}{\partial y_i} (\gamma(t)),\quad
\eta_i(t) = - \frac{\partial H}{\partial x_i}(\gamma(t)),$$ is a
solution of the VE. We can use it to reduce, using some
generalization of D'Alambert method \cite{MorRamis1, MorRamis2}
(see also \cite{MorMonograph}) our equation, obtaining the
\emph{Normal Variational Equation} (NVE), which is a linear system
of rank $2(n-1)$. In the case of $2$-degrees of freedom
hamiltonian systems, those NVE can be seen as second order linear
homogeneous differential equation.

\subsubsection{Non-integrability tools}

  Morales-Ramis theory, concerns several results that relate the
existence of first integrals of $H$ with the Galois group of the
variational equations. Along this paper we will use systematically
the following one:

\begin{thm}[\cite{MorRamis1}]\label{:MR}
  Let $H$ be a Hamiltonian in $\C^{2n}$, and $\gamma$ a particular
solution such that the {\rm NVE} has irregular singularities at
the points of $\gamma$ at infinity. Then, if $H$ is completely
integrable by terms of rational functions, then the Identity
component of Galois Group of the  {\rm NVE} is abelian.
\end{thm}

\begin{obs}
  Here, the field of coefficients of the NVE is the field
meromorphic functions on $\gamma$.
\end{obs}

\section{Some results on linear differential equations}

\subsection{Algebrization of Linear Differential Equations} For some
differential equations it is useful, if is possible, to replace the
original differential equation over a compact Riemann surface, by a
new differential equation over the Riemann sphere $\mathbb{P}^1$
(i.e., with rational coefficients) by a change of the independent
variable. This equation on $\mathbb{P}^1$ is called the {\bf
algebraic form} or {\bf algebrization} of the original equation.
Kovacic's algorithm can be applied over the algebraic form to solve
the original equation. In a more general way we will consider the
effect of a finite ramified covering on the Galois group of the
original differential equation. In \cite{MorMonograph,MorRamis1} the
following theorem is given.

\begin{thm}[Morales-Ramis \cite{MorMonograph,MorRamis1}]\label{moramis}
Let $X$ be a (connected) Riemann surface. Let $f : X'\to X)$ be a
finite ramified covering of $X$ by a Riemann surface $X'$. Let
$\nabla$ be a meromorphic connection on $X$. We set $\nabla' =
f^*\nabla$. Then we have a natural injective homomorphism $Gal
(\nabla')\to Gal (\nabla)$ of differential Galois groups which
induces an isomorphism between their Lie algebras.
\end{thm}
\begin{obs}
We observe that, in terms of the differential Galois groups, this
theorem means that the identity component of the differential
Galois group is invariant by the covering. In other words, if the
original differential equation over the Riemann surface $X'$ is
transformed by a change of the independent variable in a
differential equation over the Riemann surface $X$, then both
equations have the same identity component of the differential
Galois group.
\end{obs}

\begin{prp}[Change of the independent variable]\label{pr1}
  Let us consider the following equation, with coefficients in
$\mathbf C(x)$:
\begin{equation}\label{eq1}
  y''+a(x)y'+b(x)y=0,\quad y'={dy\over dx}
\end{equation}
and $\mathbf C(x)\hookrightarrow  L$ the corresponding
Picard-Vessiot extension. Let $(K,\delta)$ be a differential field
with $\mathbf C$ as field of constants. Let $\xi\in K$ be a
non-constat. Then, by the change of variable $x = \xi$, \eqref{eq1}
is transformed in,
\begin{equation}\label{eq2}
\ddot{y}+\left(a(\xi)\dot{\xi}-{\ddot{\xi}\over
\dot{\xi}}\right)\dot{y}+b(\xi)(\dot{\xi})^2y=0, \quad \dot z = \delta z.
\end{equation}
  Let, $K_0\subset K$ be the smallest differential field containing
$\xi$ and $\mathbf C$. Then \eqref{eq2} is a differential equation
with coefficients in $K_0$. Let $K_0\hookrightarrow L_0$ be the
corresponding Picard-Vessiot extension. Assume that
  $$\mathbf C(x) \to K_0,\quad x\mapsto \xi$$
is an algebraic extension, then
$$Gal(L_0 / K_0)^0 = Gal(L / \mathbf C(x))^0.$$
\end{prp}

\begin{dm}
By the chain rule we have
$$\frac{d}{dx} = \frac{1}{\dot\xi}\delta,$$
and,
$$\frac{d^2}{dx^2}=\frac{1}{(\dot{\xi})^2}\delta^2 - \frac{\ddot\xi}{(\dot{\xi})^3}\delta,$$
now, changing $y'$, $y''$ in \eqref{eq1} and making monic this
equation we have \eqref{eq2}
$$\ddot{y}+\left(a(\xi)\dot{\xi}-{\ddot{\xi}\over
\dot{\xi}}\right)\dot{y}+b(\xi)(\dot{\xi})^2y=0.$$ In the same
way we can obtain \eqref{eq1} through \eqref{eq2}.

  If $K_0$ is an algebraic extension of $\mathbf C(x)$, then we can
identify $K_0$ with the ring of meromorphic functions on a compact
$X$ Riemann surface, and $\xi$, is a finite ramified covering of the
Riemann sphere,
  $$X\xrightarrow{\xi} S^1.$$
If we consider $\nabla$ the meromorphic connection in $S^1$ induced
by equation \eqref{eq1}, then $\xi^*(\nabla)$, is the meromorphic
connection induced by equation \eqref{eq2} in $X$, and finally by
theorem \ref{moramis}, we conclude.
\end{dm}

 Recently Manuel Bronstein in \cite{Bronstein} has implemented an algorithm
to solve differential equation over
$\mathbf{C}(t,e^{\int_{}^{}f(t)})$ without algebrize the equation.
As immediate consequence of the proposition \eqref{eq1} we have
the following corollary.

\begin{crl}[Linear differential equation over $\mathbf{C}(t,e^{\int_{}^{}f(t)})$]Let be
$f\in\mathbf{C}(t),$
 the differential equation

\begin{equation}\label{exp}
\ddot{y}-\left(f(t)+{\dot{f}(t)\over f(t)}
-f(t)e^{\int_{}^{}f(t)}a\left(e^{\int_{}^{}f(t)}\right)\right)\dot{y}+\left(f(t)\left(e^{\int_{}^{}f(t)}\right)\right)^2b\left(e^{2\int_{}^{}f(t)}\right)y=0,
\end{equation}
by the change $x=e^{\int_{}^{}f(t)}$ is algebrizable and its
algebraic form is given by
$$y''+a(x)y'+b(x)y=0.$$
\end{crl}

\begin{obs}
In this corollary, we have the following cases.
\begin{enumerate}
\item $f=n{h'\over h}$, for a rational function $h$, $n\in\mathbf{Z}_+$, we have the trivial case, both equations
are over the Riemann sphere and they have the same differential
field, so that the equation \eqref{exp} do not need be algebrized.

\item $f={1\over n}{h'\over h}$, for a rational function $h$, $n\in\mathbf{Z}^+$, the
equation \eqref{exp} is defined over an algebraic extension of
$\mathbf{C}(t)$ and so that this equation is not necessarily over
the Riemann sphere.

\item $f\neq q{h'\over h}$, for any rational function $h$,  $q\in\mathbf{Q}$, the
equation \eqref{exp} is defined over an transcendental extension
of $\mathbf{C}(t)$ and so that the this equation is not over de
Riemann sphere.
\end{enumerate}
\end{obs}

  In first and second case, we can apply proposition \ref{pr1}, taking $K_0 = \mathbf C(t,e^{\int_{}^{}f(t)})$,
so that the identity component of the algebrized equation es
conserved. The conservation of Galois group in the third case,
requires further analysis, and will not be discussed here. We just
remark that the Galois group corresponding to original equation is
a subgroup of the Galois group of the algebrized equation.

\begin{df}[Regular and irregular singularity]\label{def1} The point $x=x_0$ is called regular singular
point (or regular singularity) of the equation \eqref{eq1} if and only if
$x=x_0$ is not an ordinary point and

$$(x-x_0)a(x),\quad (x-x_0)^2b(x),$$
are both analytic in $x=x_0.$ If $x=x_0$ is not a regular
singularity, then is an {\bf irregular singularity}.
\end{df}

\begin{obs}[Change to infinity]
To study asymptotic behaviors (or the point $x=\infty$) in
\eqref{eq1} we can take $x(t)={1\over t}$ and to analyze in
\eqref{eq2} the behavior in $t=0$. That is, to study the behavior
of the equation \eqref{eq1} in $x=\infty$ we should study the
behavior in $t=0$ of the equation
\begin{equation}\label{eq3}
\ddot{y}+\left({2\over t}-\left({1\over t^2}\right)a\left({1\over
t}\right)\right)\dot{y}+{1\over t^4}b\left({1\over t}\right)y=0.
\end{equation}
In this way, by the definition \ref{def1}, we say that $x=\infty$
is a regular singularity of the equation \eqref{eq1} if and only
if $t=0$ is a regular singularity of the equation $\eqref{eq3}$.
\end{obs}
To algebrize second order linear differential equations is easier
when the term in $\dot y$ is absent and the change of variable is
{\it{hamiltonian}}, that is, RLDE $\ddot{y}=r(t)y.$

\begin{df}[Hamiltonian change of variable]\label{def2} A change of
variable $x=x(t)$ is called hamiltonian if and only if $(x(t),\dot
x(t))$ is a solution curve of the autonomous hamiltonian system
$$H=H(x,y)={y^2\over 2}+V(x).$$
\end{df}

  Assume that we algebrize equation \eqref{eq2} through a hamiltonian
change of variables, $x = \xi(t)$. Then, $K_0 = \mathbf C(\xi, \dot\xi, \ldots)$,
but, we have the algebraic relation,
  $$(\dot\xi) ^2 = 2h - 2V(\xi), \quad h = H(\xi,\dot \xi) \in \mathbf C,$$
so that $K_0 = \mathbf C(\xi,\dot \xi)$ is an algebraic extension of $\mathbf C(x)$. We
can apply proposition \ref{pr1}, and then the identity component of the
Galois group is conserved.

\begin{prp}[Algebrization algorithm]\label{pr2}
The differential equation
$$\ddot{y}=r(t)y$$
is algebrizable through a hamiltonian change of variable $x=x(t)$
if and only if there exists $f,\alpha$ such that
$${\alpha'\over\alpha},\quad {f\over \alpha}\in \mathbf{C}(x),\text{ where } f(x(t))=r(t),\quad \alpha(x)=2(H-V(x))=\dot x^2.$$
Furthermore, the algebraic form of the equation $\ddot{y}=r(t)y$
is
\begin{equation}\label{eq4}
y''+{1\over2}{\alpha'\over \alpha}y'-{f\over\alpha}y=0.
\end{equation}
\end{prp}

\begin{dm}
Because $x=x(t)$ is a hamiltonian change of variable for the
differential equation $\ddot{y}=r(t)y$ so that $\dot x=y$, $\dot
y=\ddot x=-V'(x)$ and there exists $f,$ $\alpha$ such that
$\ddot{y}=f(x(t))y$ and $\dot x^2=2(H-V(x))=\alpha(x)$. By the
proposition \ref{pr1} we have $-f(x)=b(x)\dot{x}^2$ and $a(x)\dot
x-\ddot x/\dot x =0,$ therefore $a(x)=\ddot x/\dot x^2$ and
$b(x)=-f(x)/\alpha(x)$. In this way $\ddot y=r(t)y$ is
algebrizable if and only if $a(x), b(x) \in\mathbf{C}(x).$ As
$2(H-V(x))=\alpha(x)$ then $\alpha'(x)=-2V'(x)=2\ddot x,$ and
therefore $a(x)={1\over 2}{\alpha'(x)\over \alpha (x)}.$ In this
way, we obtain the equation \eqref{eq4}.
\end{dm}
As consequence of the proposition \ref{pr2} we have the following
result.
\begin{crl}\label{cor1}
Let be $r(t)=g(x_1,\cdots, x_n)$, where $x_i=e^{\lambda_i t}$,
$\lambda_i\in\mathbf{C}$. The differential equation
$\ddot{y}=r(t)y$ is algebrizable if and only if
$${\lambda_i\over \lambda_j}\in \mathbf{Q},\quad 1\leq i\neq j\leq n,\quad g\in \mathbf{C}(x).$$
Furthermore, we have $\lambda_i=c_i\lambda$, where
$\lambda\in\mathbf{C}$ and $c_i\in \mathbf{Q}$ and one change of
variable is
\begin{displaymath}
x=e^{\lambda t\over q},\text{ where } c_i={p_i\over q_i}, \text{
}\gcd(p_i,q_i)=1 \text{ and } q=mcm(q_1,\cdots,q_n).
\end{displaymath}
\end{crl}

\begin{obs}[Using the algebrization algorithm]
To algebrize the differential equation $\ddot y=r(t)y$ it should
keep in mind the following steps.

\begin{description}
\item[Step 1] Find a hamiltonian change of variable $x=x(t)$.
\item[Step 2] Find $f$ and $\alpha$ such that $r(t)=f(x(t))$ and
$(\dot x (t))^2=\alpha(x(t))$.
\item[Step 3] Write $f(x)$ and $\alpha(x)$.
\item[Step 4] Verify if $f(x)/\alpha (x)\in\mathbf{C}(x)$ and $\alpha'(x)/\alpha
(x)\in\mathbf{C}(x)$ to see if the RLDE is algebrizable or not.
\item[Step 5] If the RLDE is algebrizable, write the algebraic
form of the original equation such as follows
$$y''+{1\over2}{\alpha'\over \alpha}y'-{f\over\alpha}y=0.$$
\end{description}
When we have algebrized second order linear differential equation
we study its integrability and its Galois groups.
\end{obs}

\begin{obs}[Monic Polynomials]\label{monic}
Let be the reduced linear differential equation (RLDE)
$$\ddot y=\left(\sum_{k=0}^nc_kt^k\right)y,\quad c_k\in\mathbf{C}, \quad k=1,\cdots,n.$$
By the algebrization algorithm we can take $x=\mu t$,
$\mu\in\mathbf{C}$, so that
$$\dot x=\mu,\quad \alpha(x)=\mu^2,\quad \alpha'(x)=0 \text{ and } f(x)=\sum_{k=0}^nc_k\left({x\over \mu}\right)^k,\quad c_k,\mu\in\mathbf{C}.$$
Now, by the equation \eqref{eq4} the new differential equation is
$$y''=\left(\sum_{k=0}^n\left({c_k\over \mu^{k+2}}\right)x^k\right)y,\quad c_k,\mu\in\mathbf{C}.$$

In general, for $\mu=\sqrt[n+2]{c_{n}}$ we can obtain the equation

$$y''=\left(x^n+\sum_{k=0}^{n-1}d_kx^k\right)y,\quad d_k=\left({c_k\over \mu^{k+2}}\right),\quad k=0,\cdots, n-1.$$

Furthermore, we can observe, by definition \ref{def1} and equation
\eqref{eq3}, that the point at $\infty$ is an irregular
singularity for the differential equation with non-constant
polynomial coefficients because using equation \eqref{eq3} we can
see that zero is not an ordinary point and neither is a regular
singularity for the differential equation.
\end{obs}

\begin{obs}[Extended Mathieu]\label{Mathiewobs}
The Extended Mathieu differential equation is
\begin{equation}\label{Mathew}
\ddot{y}=(a+b\sin t+c\cos t)y,
\end{equation}

in particular, when $b=0$ or $c=0$ and $|a|+|b|\neq 0,$ we have
the so called Mathieu equation. Applying the corollary \ref{cor1}
and the steps of the algorithm we have $x=e^{it}$, $\dot x=ix,$
therefore
$$f(x)={(b+c)x^2+2ax+c-b\over 2x},\quad \alpha(x)=-x^2,\quad \alpha'(x)=-2x,$$

so that the algebraic form of the equation \eqref{Mathew} is
\begin{equation}\label{AlMat}
y''+{1\over x}y'+{(b+c)x^2+2ax+c-b\over 2x^3}y=0,
\end{equation}
Making the change $x=1/z$ in the equation \eqref{AlMat} we obtain
\begin{equation}\label{matirr}
\ddot{\zeta}+\left({1\over
z}\right)\dot{\zeta}+\left({(c-b)z^2+2az+(b+c)\over
2z^3}\right)\zeta=0.
\end{equation}
We can observe, by definition \ref{def1} and equation \eqref{eq3},
that $z=0$ is an irregular singularity for the equation
\eqref{matirr} and therefore $x=\infty$ is an irregular
singularity for the equations \eqref{AlMat} and \eqref{Mathew}.

Now, we compute the Galois group and the integrability in equation
\eqref{AlMat}. So that, the RLDE is given by
\begin{equation}\label{matred}
\xi''=-\left({(b+c)x^2+(2a+1)x+c-b\over 2x^3}\right)\xi.
\end{equation}
Applying the Kovacic's algorithm, see Appendix A, we can see that
for $b\neq -c$ this equation falls in case 2: $(c_3,\infty_3),$
$E_0=\{3\}$, $E_{\infty}=\{1\}$ and so that $D=\emptyset$ because
$m=-1\notin \mathbf{Z}_+$. In this way we have that the equation
\eqref{matred} is not integrable, the Galois Group is the
connected group $SL(2,\mathbf{C}),$ and finally, by the theorem
\ref{moramis} the identity component of the Galois group for
equation \eqref{Mathew} is exactly $SL(2,\mathbf{C})$ that is not
an abelian group. In the same way, for $b=-c$ we have the
equations
$$\ddot{y}=(a-be^{-it})y,\quad y''+{1\over x}y'+{2ax-2b\over 2x^3}y=0,$$
in which $\infty$ continues being an irregular singularity. Now,
Its RLDE is given by
$$\xi''=-\left({(2a+1)x-2b\over 2x^3}\right)\xi,$$
applying Kovacic's algorithm we can see that this equation falls
in case 2: $(c_3,\infty_2),$ $E_0=\{3\}$, $E_{\infty}=\{0,2,4\}$,
so that $D=\emptyset$ because $1/2(e_{\infty}-e_0)\notin
\mathbf{Z}_+.$ This means that the Galois group continues being
$SL(2,\mathbf{C}).$ Using this result, taking $\epsilon$ instead
of $i$, we can say that in the case of {\it{harmonic oscillator
with exponential waste}}
$$\ddot{y}=(a+be^{-\epsilon t})y, \quad \epsilon>0,$$
the point at $\infty$ is an irregular singularity and the identity
component of the Galois group is $SL(2,\mathbf{C})$.
\end{obs}

\subsection{Galois groups of Schr\"odinger equations with polynomial potential}\label{:sspoly}

Let be the reduced linear differential equation (RLDE)
$$\ddot \xi=\left(\sum_{k=0}^nc_kt^k\right)\xi,\quad a_k\in\mathbf{C}.$$

By the remark \ref{monic}, through the change of variable $x=\mu t,$
$\mu=\sqrt[n+2]{c_k},$ this equation become in

$$\xi''=P_n(x)\xi,\quad P_n(x)\text{ is a monic polynomial of degree }n.$$

Kovacic in \cite{Kov} remarked that for $n=2k+1$ there are not
liouvillian solutions and therefore the Galois group of the RLDE is
$SL(2,\mathbf{C})$. Using the Kovacic's algorithm (see Appendix), we
can see that for $n=2k$ the equation falls in case 1, specifically
in $c_0$ (because has not poles) and $\infty_3$ (because $\circ
r_{\infty}=-2k$), that is $\{c_0,\infty_3\}$.

\begin{lm}[Completing Squares]\label{cosq}
Every monic polynomial of degree even can be written in one only way
completing squares, that is
$$Q_{2n}(x)=x^{2n}+\sum_{k=0}^{2n-1}q_kx^k=\left(x^n+\sum_{k=0}^{n-1}a_kx^k\right)^2+\sum_{k=0}^{n-1}b_kx^k.$$
\end{lm}

\begin{dm}
Firstly, we can see that
$$\left(x^n+\sum_{k=0}^{n-1}a_kx^k\right)^2=x^{2n}+2\sum_{k=0}^{n-1}a_kx^{n+k}+\sum_{k=0}^{n-1}\sum_{j=0}^{n-1}a_ka_jx^{k+j},$$
so that by indeterminate coefficients we have

$$a_{n-1}={q_{2n-1}\over 2},\quad a_{n-2}={q_{2n-2}-a^2_{n-1}\over 2},\quad a_{n-3}={q_{2n-3}-2a_{n-1}a_{n-2}\over 2},\cdots,$$

$$a_0={q_n-2a_1a_{n-1}-2a_2a_{n-2}-\cdots\over 2},\quad
b_0=q_0-a_0^2,\quad b_1=q_1-2a_0a_1,\quad \cdots,$$
$$b_{n-1}=q_{n-1}-2a_0a_{n-1}-2a_1a_{n-2}-\cdots,$$
therefore, we has proven the lemma.
\end{dm}

By case 1, $\{c_0,\infty_3\}$, of the kovacic's algorithm, remark
\ref{rkov2} [I5] (See Appendix), remark \ref{monic} and by lemma
\ref{cosq} we have proven the following theorem.

\begin{thm}[Galois groups in polynomial case]\label{polynint}
Let us consider the equation,
$$\ddot \xi = Q(x)\xi,$$
with $Q(x)$ a polynomial of degree $k>0$. Then, it falls in one of
the following cases:
\begin{enumerate}
\item $k=2n$ is even, $\pm b_{n-1}-n=2m,$ $m\in \mathbf{Z}_+$, and there
exist a monic polynomial $P_m$ of degree $m$ satisfying
\begin{displaymath}
P_m'' + 2\left(x^n+\sum_{k=0}^{n-1}a_kx^k\right)P_m' +
\left(nx^{n-1}+\sum_{k=0}^{n-2}(k+1)a_{k+1}x^{k} +
\sum_{k=0}^{n-1}b_kx^k\right)P_m = 0,
\end{displaymath}
the solutions are given by
\begin{eqnarray*}
\xi_1&=&P_me^{{x^{n+1}\over n+1}+\sum_{k=0}^{n-1}{a_kx^{k+1}\over k+1}}, \\
\quad \xi_2&=&P_me^{{x^{n+1}\over
n+1}+\sum_{k=0}^{n-1}{a_kx^{k+1}\over k+1}}
  \int_{}^{}{dx\over P_m^2e^{2\left({{x^{n+1}\over n+1}+\sum_{k=0}^{n-1}{a_kx^{k+1}\over k+1}}\right)}}.
\end{eqnarray*}
and the Galois group is $\mathbf{C}^*\ltimes\mathbf{C}$
(non-abelian, resoluble, connected group).
\item
The equation has not liouvillian solutions and its Galois group is
$SL(2,\mathbf{C})$.
\end{enumerate}
\end{thm}

\begin{obs}[Quadratic case]\label{:rkm2}

consider the case where $r$ is a polynomial of degree two:
$$\ddot y = \left(At^2 + Bt+C\right)y.$$
There are no poles and the order at $\infty$, $\circ r_{\infty}$, is
$-2$, so we need to follow case 1 in $\{c_0,\infty_3\}$ of the
algorithm. Now, by remark \ref{monic} and by lemma \ref{cosq} we
have
$$y''=\left((x+a)^2+b\right)y.$$
We find that
\begin{align*}
    [\sqrt{r}]_{\infty}&= x + a \\
    \alpha_\infty^\pm &= {1\over 2}\left(\pm b - 1\right)\\
    m &= \alpha_\infty^+\quad\text{or}\quad\alpha_\infty^-.
\end{align*}
If $b$ is not an odd integer then $m$ cannot be an integer so case 1
cannot hold so the RLDE in $x$ has no Liouvillian solutions. If $b$
is an odd integer than we can complete steps 2 and 3 and actually
(only) find a solution it which is
$$y=P_me^{{x^2\over 2}+ax}.$$

In particular, the quantum harmonic oscillator
$$y''=(x^2+\lambda)y$$
is integrable when $\lambda$ is an odd integer and the only one
solution obtained by means of kovacic's algorithm is given by
$$y=H_me^{x^2\over 2},$$
where $H_m$ denotes the classical Hermite's polynomials.
\end{obs}

  For another approach to this problem see Vidunas \cite{Vidunas} and Zoladek in \cite{Zoladek}.

\section{Determining families of Hamiltonians with specific NVE}

  Let us consider a two degrees of freedom classical hamiltonian,
$$H = \frac{y_1^2+y_2^2}{2} + V(x_1,x_2).$$
  $V$ is the \emph{potential function}, and it is assumed to be
analytical in some open subset of $\C^2$. The evolution of the
system is determined by Hamilton equations:

$$ \dot x_1 = y_1,\quad \dot x_2 = y_2, \quad \dot y_1 =
-\frac{\partial V}{\partial x_1},\quad \dot y_2 = -\frac{\partial
V}{\partial x_2}.$$

  Let us assume that the plane $\Gamma = \{x_2=0, y_2 = 0\}$ is
an invariant manifold of the hamiltonian. We keep in mind that the
family of integral curves lying on $\Gamma$ is parameterized by
the energy $h = H|_\Gamma$, but we do not need to use it
explicitly. We are interested in studying the linear approximation
of the system near $\Gamma$. Since $\Gamma$ is an invariant
manifold, we have
$$\left.\frac{\partial V}{\partial x_2}\right|_\Gamma = 0,$$
so that the Normal Variational Equation for a particular solution
$$t\mapsto\gamma(t) =(x_1(t), y_1 = \dot x_1(t), x_2 = 0, y_2 = 0),$$
is written,
$$\dot \xi = \eta,\quad \dot \eta = -\left[\frac{\partial^2 V}{\partial x_2^2}(x_1(t),0)\right] \xi.$$

Let us define,
$$\phi(x_1) = V(x_1, 0),\quad \alpha(x_1) =
- \frac{\partial^2 V}{\partial x_2^2}(x_1,0),$$ and then we write
the second order Taylor series in $x_2$ for $V$, obtaining the
following expression for $H$
\begin{equation}\label{:Hgeneral}
  H = \frac{y_1^2+y_2^2}{2} +  \phi(x_1) - \alpha(x_1)\frac{x_2^2}{2}
  + \beta(x_1,x_2)x_2^3,
\end{equation}
which is the \emph{general form of a classical a Hamiltonian, with
invariant plane $\Gamma$}.The NVE associated to any integral curve
lying on $\Gamma$ is,
\begin{equation}\label{:NVE}
\ddot \xi = \alpha(x_1(t)) \xi.
\end{equation}

\subsection{General Method}

  We are interested in compute Hamiltonians of the family
(\ref{:Hgeneral}), such that its NVE (\ref{:NVE}) belongs to a
specific family of Linear Differential Equations. Then we can apply
our results about the integrability of this LDE, and Morales-Ramis
theorem to obtain information about the non-integrability of such
Hamiltonians.

  From now on, we will write $a(t) = \alpha(x_1(t))$, for a generic
curve $\gamma$ lying on $\Gamma$, parameterized by $t$. Then, the
NVE is written
\begin{equation}\label{:NVEa}
\ddot \xi = a(t)\xi.
\end{equation}

  {\bf Problem. }\emph{Assume that $a(t)$ is a root of a \underline{given differential
polynomial} $Q(a, \dot a, \ddot a, \ldots ) \in \C[a, \dot a,
\ddot a,\ldots]$. We want to compute all hamiltonians in
\eqref{:Hgeneral} satisfying such a condition.}

\medskip

In this section we give a method to compute, for any given $Q(a,\dot a,\ldots)$,
the family of classical hamiltonians with invariant plane $\Gamma$ such that,
for any integral curve lying on $\Gamma$, the coefficient $a(t)$ of the NVE satisfies,
\begin{equation}\label{:Q}
Q(a,\dot a,\ddot a,\ldots) =0,
\end{equation}
by solving certain differential equations. This method lies under the
calculus done by J. Morales and C. Sim\'o in \cite{MorSimo}.

\medskip

  We should notice that, for a generic integral curve
$\gamma(t) = (x_1(t),y_1 = \dot x_1(t))$ lying on $\Gamma$,
equation (\ref{:NVEa}) depends only of the values of functions $\alpha$, and $\phi$.
It depends of $\alpha(x_1)$, since $a(t) = \alpha(x_1(t))$.
We observe that the curve $\gamma(t)$ is a solution of the
restricted Hamiltonian,
\begin{equation}\label{:restrictedH}
  h = \frac{y_1^2}{2} + \phi(x_1)
\end{equation}
  whose associated Hamiltonian vector field is,
\begin{equation}\label{:Xh}
  X_h = y_1\frac{\partial}{\partial x_1} - \frac{d\phi}{d x_1}\frac{\partial}{\partial y_1},
\end{equation}
 thus $x_1(t)$ is a solution of the differential equation, $\ddot x_1 = -\frac{d\phi}{d x_1},$
and then, the relation of $x_1(t)$ is given by $\phi$.

\medskip

 Since $\gamma(t)$ is an integral curve of $X_h$, for any function
$f(x_1,y_1)$ defined in $\Gamma$ we have
  $$\frac{d}{d t}\gamma^*(f) = \gamma^*(X_h f),$$
where $\gamma^*$ denote the usual pullback of functions. Then,
using $a(t) = \gamma^*(\alpha)$, we have for each $k\geq 0$,
\begin{equation}
\frac{d^k a}{d t^k} = \gamma^*(X_h^k\alpha),
\end{equation}
so that,
$$Q(a,\dot a, \ddot a,\ldots ) = Q(\gamma^*(\alpha),\gamma^*(X_h\alpha),\gamma^*(X_h^2\alpha),\ldots).$$

  There is an integral curve of the Hamiltonian passing through each point of $\Gamma$, so that
we have proven the following.

\begin{prp}\label{:prop2}
  Let $H$ be a Hamiltonian of the family {\rm (\ref{:Hgeneral})}, and $Q(a,\dot a,\ddot a,\ldots)$ a differential
polynomial with constants coefficients. Then, for each integral
curve lying on $\Gamma$, the coefficient $a(t)$ of the NVE {\rm
(\ref{:NVEa})} verifies $Q(a,\dot a, \ddot a, \ldots,) = 0$, if
and only if the function
$$ \hat Q(x_1,y_1) = Q(\alpha, X_h\alpha, X_h^2 \alpha, \ldots),$$
vanish on $\Gamma$.
\end{prp}

\begin{obs}
  In fact the NVE of a integral curve depends on the parameterization.
Our criterion does not depend on any choice of parameterization of the integral curves.
This is simple, the NVE corresponding to different
parameterizations of the same integral curve are related by a
translation of time $t$. We need just observe that a polynomial
$Q(a,\dot a, \ddot a,\ldots)$ with constant coefficients is
invariant of the group by translations of time. Then if the
coefficient $a(t)$ of the NVE (\ref{:NVEa}) for certain
parameterization of an integral curve $\gamma(t)$ satisfied
$\{Q=0\}$, then it is also satisfied for any other right
parameterization of the curve.
\end{obs}

  Next, we will see that $\hat Q(x_1,y_1)$ is a polynomial in $y_1$ and its coefficients are
differential polynomials in $\alpha, \phi$. If we write down the expressions for successive Lie derivatives
of $\alpha$, we obtain
\begin{equation}\label{:armonico}
X_h \alpha = y_1\frac{d \alpha}{d x_1},
\end{equation}

\begin{equation}\label{:airy}
X_h^2 \alpha = y_1^2\frac{d^2\alpha}{d x_1^2} - \frac{d \phi}{d x_1}\frac{d \alpha}{ d x_i}
\end{equation}

\begin{equation}\label{:qarmonic}
X_h^3\alpha = y_1^3\frac{d^3\alpha}{dx_1^3} - y_1\left(
    \frac{d}{dx_1}\left(\frac{d\phi}{dx_1}\frac{d\alpha}{dx_1}\right)
    + 2\frac{d\phi}{dx_1}\frac{d^2\alpha}{dx_1^2}\right)
\end{equation}

$$X_h^4\alpha = y_1^4\frac{d^4\alpha}{dx_1^4}-y_1^2\left(\frac{d}{dx_1}\left(
    \frac{d}{dx_1}\left(\frac{d\phi}{dx_1}\frac{d\alpha}{dx_1}\right) + 2\frac{d\phi}{dx_1}\frac{d^2\alpha}{dx_1^2}\right)
    +3\frac{d^3\alpha}{dx_1^3}\frac{d\phi}{dx_1}\right) +$$

\begin{equation}
+\left(
    \frac{d}{dx_1}\left(\frac{d\phi}{dx_1}\frac{d\alpha}{dx_1}\right) + 2\frac{d\phi}{dx_1}\frac{d^2\alpha}{dx_1^2}\right)
    \frac{d\phi}{dx_1}.
\end{equation}

  In general form we have,
\begin{equation}\label{:induction}
X_h^{n+1}\alpha = y_1\frac{\partial X_h^{n}\alpha}{d x_1} - \frac{d\phi}{d x_1}\frac{\partial X_h^{n}}{\partial y_1},
\end{equation}
it inductively follows that they all are polynomial in $y_1$ with coefficients differential polynomials in
$\alpha,\phi$.
If we write it down explicitly,
\begin{equation}\label{:Ex}
X_h^n\alpha = \sum_{n\geq k \geq 0} E_{n,k}(\alpha,\phi)y_1^k
\end{equation}
we can see that the coefficients $E_{n,k}(\alpha,\phi) \in
\C\left[\alpha,\phi,\frac{d^r\alpha}{d
x_1^r},\frac{d^s\phi}{dx_1^s}\right]$, satisfies the following
recurrence law,
\begin{equation}\label{:rlaw}
  E_{n+1,k}(\alpha,\phi) = \frac{d}{d x_1} E_{n,k-1}(\alpha,\phi) - (k+1)E_{n,k+1}(\alpha,\phi)\frac{d\phi}{d x_1}
\end{equation}
with initial conditions,
\begin{equation}\label{:condition}
E_{1,1}(\alpha,\phi) = \frac{d \alpha}{d x_1},\quad E_{1,k}(\alpha,\phi) = 0 \,\,\,\forall k\neq 1.
\end{equation}

\begin{obs}\label{:remark}
  The recurrence law (\ref{:rlaw}) and initial conditions (\ref{:condition}) determine the coefficients
$E_{n,k}(\alpha,\phi)$. We can compute the value of some of them easily:
\begin{itemize}
\item $E_{n,n}(\alpha,\phi) = \frac{d^n\alpha}{d x_1^n}$ for all $n \geq 1$.
\item $E_{n,k}(\alpha,\phi)= 0$ if $n-k$ is odd, or $k<0$, or $k>n$.
\end{itemize}
\end{obs}

\subsection{Some Examples}

  Here we compute families of hamiltonians (\ref{:Hgeneral}) such give rise to specific
NVE. Although, in order to this computations, we need to solve Polynomial Differential Equations,
we will see that we can deal with this in a branch of cases. Particularly, when $Q$ is a Differential
Linear Operator, we will obtain equations that involve products of few Linear Differential Operator.

\begin{ej}
  \emph{Harmonic oscillator} equation is
\begin{equation}\label{:eqarmonico}
\ddot \xi = c_0\xi,
\end{equation}
with $c_0$ constant. Then, a hamiltonian of type (\ref{:Hgeneral}) gives such NVE
if $\dot a = 0$. Looking
at formula (\ref{:armonico}), it follows that $\frac{d \alpha}{d x_1}=0$, so that $\alpha$ is
a constant.
We conclude that the general form of a Hamiltonian (\ref{:Hgeneral}) which give
rise to NVE of the type (\ref{:eqarmonico}) is,
$$H = \frac{y_1^2+y_2^2}{2} + \phi(x_1) + \lambda_0x_2^2 + \beta(x_1,x_2)x_2^3,$$
being $\lambda_0$ a constant, and $\phi,\beta$ arbitrary analytical functions.
\end{ej}

\begin{ej}
In \cite{Audin}, M. Audin notice that the hamiltonian,
$$\frac{y_1^2+y_2^2}{2} + x_1x_2^2$$
gives an example of a simple non-integrable classical hamiltonian, since its NVE along any
integral curve in $\Gamma$ is an \emph{Airy equation}. Here we compute the family of classical
hamiltonians that have NVE of type Airy for integral curves lying on $\Gamma$. General form of Airy
equation is
\begin{equation}\label{:eqairy}
\ddot\xi = (c_0 + c_1 t)\xi
\end{equation}
with $c_0,c_1\neq 0$ two constants. If follows that a hamiltonian
gives rise to NVE of this type if $\ddot a = 0$, and $\dot a \neq
0$. The equation $\ddot a = 0$ gives, by proposition \ref{:prop2}
as we see in formula (\ref{:airy}), the following system:
\begin{equation}
 \frac{d^2\alpha}{d x_1^2} = 0,\quad \frac{d\phi}{d x_1}\frac{d \alpha}{d x_1} = 0.
\end{equation}
It split in two independent systems,
\begin{equation}
\frac{d\alpha}{dx_1} = 0,\quad \left\{\begin{array}{c}
\frac{d^2\alpha}{dx_1^2} = 0 \\
\frac{d\phi}{dx_1} = 0
 \end{array}\right.
\end{equation}
Solutions of the first one fall into the previous case of
\emph{harmonic oscillator}. Then, taking the general solution of
the second system, we conclude that the general form of a
classical hamiltonian of type (\ref{:Hgeneral}) with Airy NVE is:
\begin{equation}
H = \frac{y_1^2+y_2^2}{2}+\lambda_0 + \lambda_1 x_2^2 + \lambda_2 x_1x_2^2 + \beta(x_1,x_2)x_2^3,
\end{equation}
with $\lambda_2 \neq 0$.
\end{ej}

\subsubsection{NVE quantum harmonic oscillator}

  Let us consider now equations with $\frac{d^3 a}{dt^3}=0$, and $\frac{d^2 a}{dt^2}\neq 0$,
it is
\begin{equation}\label{:quadratic}
\ddot \xi = (c_0 + c_1 t + c_2 t^2)\xi
\end{equation}
with $c_2\neq 0$. Those equation can be reduced to a
\emph{quantum harmonic oscillator equation} by an affine change of
$t$, and its integrability has been studied using Kovacic's
Algorithm. Using proposition \ref{:prop2} and formula
(\ref{:qarmonic}), we obtain the following system of differential
equations for $\alpha$ and $\phi$:
$$\frac{d^3\alpha}{dx_1^3}=0,\quad \frac{d^\alpha}{d x_1}\frac{d^2\phi}{dx_1^2}+ 3\frac{d^2\alpha}{d x_1^2}
\frac{d\phi}{dx_1}=0.$$
General solution of the first equation is
$$\alpha = \frac{\lambda_1}{2} + \frac{\lambda_2}{2}x_1 + \frac{\lambda_3}{2}x_1^2,$$
and substituting it into the second equation we obtain a Linear Differential Equation for $\phi$,
$$\frac{d^2\phi}{dx_1^2} + 3\frac{2\lambda_3}{\lambda_2+2\lambda_3 x_1}\frac{d\phi}{d x_1} = 0,$$
this equation is integrated by two quadratures, and its general solution is
$$\phi = \frac{\lambda_4}{(\lambda_2 + 2\lambda_3 x_1)^2} + \lambda_0.$$
We conclude that the general formula for hamiltonians of type (\ref{:Hgeneral}) with NVE
(\ref{:quadratic}) for any integral curve lying on $\Gamma$ is
$$H = \frac{y_1^2+y_2^2}{2} + $$
\begin{equation}\label{:Hqarmonic}
\frac{\lambda_4}{(\lambda_2 + 2\lambda_3 x_1)^2} + \lambda_0 -
\lambda_1 x_2^2 - \lambda_2 x_1x_2^2 - \lambda_3 x_1^2 x_2^2 +
\beta(x_1,x_2)x_2^3,
\end{equation}
with $\lambda_3\neq 0$.

\begin{obs}
  Formula (\ref{:Hqarmonic}) is the first example, in this paper, in which we find non-linear
dynamics in the invariant plane $\Gamma$. Notice that this dynamic
is continuously deformed to linear dynamics when $\lambda_4$ tends
to zero. In general case, for a fixed energy $h$, we have the general integral
of the equation:
$$8\lambda_3^2h^2(t-t_0)^2 = h(\lambda_2 + 2\lambda_3x_1)^2 - \lambda_4.$$
\end{obs}

\subsubsection{NVE with polynomial coefficient $a(t)$ of odd degree}

  Let us consider for $n>0$ the following differential polynomial, $$Q_m(a,\dot a,\ldots)  = \frac{d^m a}{dt^m}.$$
It is obvious that $a(t)$ is polynomial of degree $n$ if and only if $Q_n(a,\dot a,\ldots)\neq 0$ and
$Q_{n+1}(a,\dot a,\ldots) = 0$.

  Looking a proposition \ref{:prop2}, we see that a hamiltonian (\ref{:Hgeneral}) has NVE along a generic integral
curves lying on $\Gamma$,
\begin{equation}\label{:epol}
\ddot \xi = P_n(t)\xi, \end{equation}
with $P_n(t)$ of degree $n$,
if and only if $X_h^n\alpha\neq 0$ and $X^{n+1}_h\alpha$ vanish on
$\Gamma$. Let us remember expression (\ref{:Ex}), $X^{n+1}_h\alpha$ vanish in $\Gamma$ if and only if
$(\alpha,\phi)$ is a solution of the differential system,
  $$R_{n+1} = \{E_{n+1,0}(\alpha,\phi)=0,\ldots, E_{n+1,n+1}(\alpha,\phi)=0\}.$$
Using the recurrence law (\ref{:rlaw}) defining the differential polynomials $E_{n,k}(\alpha,\phi)$ we are
going to compute the family of hamiltonians giving rise to $a(t)$ polynomial of odd degree.

\begin{lm}
  Let $(\alpha,\phi)$ be a solution of $R_{2m}$. Then, if $\frac{d\phi}{dx_1}\neq 0$, then $(\alpha,\phi)$ is a solution
of $R_{2m-1}$.
\end{lm}

\begin{dm}
By remark \ref{:remark} $E_{2m-1,2k}(\alpha,\phi) = 0$ for all $m-1\geq k\geq 0$. Then let us proof that
$E_{2m-1,2k+1}(\alpha,\phi)$ for all $m-2\geq k\geq 0$.

In the first step of the recurrence law defining $R_{2m}$,
  $$0 = E_{2m,0}(\alpha,\phi) = \frac{dE_{2m-1,1}}{dx_1}(\alpha,\phi) - \frac{d\phi}{d x_1}E_{2m-1,1}(\alpha,\phi),$$
we use $\frac{d\phi}{dx_1}\neq  0$, and remark \ref{:remark},
$E_{2m-1,-1}(\alpha,\phi) = 0$ to obtain,
  $$E_{2m-1,1}(\alpha,\phi) = 0.$$
If we assume $E_{2m-1, 2k+1}(\phi,\alpha) = 0$, substituting it in the recurrence law
  $$E_{2m,2k+1}(\alpha,\phi) = \frac{d E_{2m-1,2k}}{dx_1}(\alpha,\phi) -
2(k+1)\frac{d\phi}{dx_1}E_{2m-1,2(k+1)}(\alpha,\phi),$$
we obtain that
  $$E_{2m-1,2(k+1)}(\alpha,\phi)=0,$$
and we conclude by finite induction.
\end{dm}

\begin{crl}
  Let $H$ be a classical hamiltonian of type {\rm (\ref{:Hgeneral})}, then the following statements are equivalent,
\begin{enumerate}
\item The NVE for generic integral curve {\rm (\ref{:NVEa})} lying on $\Gamma$ has polynomial coefficient $a(t)$
of degree $2m-1$.
\item  $H$ is written,
\begin{equation}\label{:odd} H = \frac{y_1^2+y_2^2}{2} + \lambda_0 - P_{2m-1}(x_1)x_2^2 +
\beta(x_1,x_2)x_2^3,\end{equation} for $\lambda_0$ constant, and
$P_{2m-1}(x_1)$ polynomial of degree $2m-1$.
\end{enumerate}
\end{crl}

\begin{dm}
  It is clear that condition \emph{1.} is satisfied if and only if $(\alpha,\phi)$ is a solution of $R_{2m}$ and
it is not a solution of $R_{2m-1}$. By the previous lemma, it implies $\frac{d\phi}{dx_1}= 0$, and then
the system $R_{2m}$ is reduced to $\frac{d^{2m}\alpha}{dx_1^{2m}}$ and then, $\phi$ is a constant and $\alpha$ is
a polynomial of degree at most $2m-1$.
\end{dm}

\subsubsection{NVE Mathieu extended}
  This is the standard Mathieu equation,
  \begin{equation}\label{:Mathieu}
  \ddot \xi =(c_0 + c_1\cos(\omega t))\xi, \quad \omega \neq 0.
  \end{equation}
  We can not apply our method to compute the family of
hamiltonians corresponding to this equation, because
$\{c_0+c_1\cos(\omega t)\}$ is not the general solution of any differential
polynomial with constant coefficients. But, let us consider
\begin{equation}\label{:Mequation}
 Q(a) = \frac{d^3a}{dt^3}+\omega^2\frac{da}{dt},
\end{equation}
the general solution of $\{Q(a)=0\}$ is
$$a(t) = c_0+c_1\cos(\omega t)+c_2\sin(\omega t).$$
Just notice that,
$$c_1\cos(\omega t)+c_2\sin(\omega t) = \sqrt{c_1^2+c_2^2}
\cos\left(\omega t+\arctan\frac{c_2}{c_1}\right),$$
thus NVE \eqref{:NVEa}, when $a$ is a solution of \eqref{:Mequation}, is
reducible to Mathieu equation (\ref{:Mathieu}) by a translation
of time.

\medskip

Using Proposition \ref{:prop2}, we find the system of differential equations
that determine the family of hamiltonians,
$$\frac{d^3\alpha}{dx_1^3}=0,\quad
\frac{d\alpha}{dx_1}\frac{d^2\phi}{dx_1^2}+3\frac{d^2\alpha}{dx_1}\frac{d\phi}{dx_1}-\omega^2\frac{d\alpha}{dx_1}=0.$$
General solution of the first equation is
$$\alpha = \lambda_0 + \lambda_1x_1 + \lambda_2x_1^2.$$
substituting it in the second equation, and writing $y =
\frac{d\phi}{dx_1}$, we obtain a non homogeneous linear differential
equation for $y$,
\begin{equation}\label{:yeq}
\frac{dy}{dx_1}+\frac{6\lambda_2y}{\lambda_1+2\lambda_2x_1}=\omega^2.
\end{equation}
We must distinguish two cases depending on the parameter. If
$\lambda_2 = 0$, then we just integrate the equation by trivial
quadratures, obtaining
$$\phi = \mu_0 + \mu_1x_1 + \frac{\omega^2x_1^2}{2}$$
and then,
\begin{equation}\label{:Mhamiltonian1}
H = \frac{y_1^2+y_2^2}{2}+\mu_0 + \mu_1x_1 +\frac{\omega^2
x_1^2}{2}- \lambda_0x_2^2 - \lambda_1x_1x_2^2 + \beta(x_1,x_2)x_2^3,
\end{equation}
If $\lambda_2\neq 0$, then we can reduce the equation to separable
using
$$u = \frac{6\lambda_2y}{\lambda_1+2\lambda_2x_1},$$
obtaining
$$\frac{3du}{3\omega^2-4u} = \frac{6\lambda_2 dx}{\lambda_1+2\lambda_2 x_1},\quad u = \frac{3\omega^2}{4} + \frac{3 \mu_1}{4(\lambda_1+2\lambda_2x_1)^4},$$
and then
$$ y = \frac{1}{8\lambda_2}\left(\omega^2\lambda_1 + 2\omega^2\lambda_2x_1 + \frac{\mu_1}{(\lambda_1 + 2\lambda_2x_1)^3}\right),$$
and finally we integrate it to obtain $\phi$,
$$\phi = \int y dx_1 = \mu_0 -
\frac{\mu_1}{32\lambda_2^2}\frac{1}{(\lambda_1+2\lambda_2x_1)^2}+\frac{\omega^2\lambda_1x_1}{8\lambda_2}+\frac{\omega^2x_1^2}{8},$$
scaling the parameters adequately we write down the general formula
for the hamiltonian,
$$ H = \frac{y_1^2+y_2^2}{2} + \mu_0 +
\frac{\mu_1}{(\lambda_1+2\lambda_2x_1)^2}+\frac{\lambda_1\omega^2 x_1}{8\lambda_2}
+ \frac{\omega^2 x_1^2}{8} + $$
\begin{equation}\label{:Mhamiltonian2}
 -\lambda_0 x_2^2 -  \lambda_1x_1x_2^2 - \lambda_2x_1^2x_2^2 +
\beta(x_1,x_2)x_2^3.
\end{equation}

\subsection{Application of non-integrability criteria}

\subsubsection{NVE with $a(t)$ polynomial}

  According to  theorem \ref{polynint} (see Appendix), the Galois group corresponding to
equations,

 $$\ddot \xi = P(t)\xi$$

  where $P(t)$, is a non-constant polynomial, is a connected non-abelian group.
So that, we can apply theorem \ref{:MR}, and we get the following result.

\begin{prp}
Hamiltonians,
$$H = \frac{y_1^2+y_2^2}{2} + x^2Q(x_1) + \beta(x_1,x_2)x_1^2$$
where $Q(x_1)$ is a non-constant polynomial, and $\beta(x_1,x_2)$
analytic function around $\Gamma$, do no admit any additional
rational first integral.
\end{prp}

\begin{crl}
Every integrable (by rational functions) polynomial potential
with invariant plane $\Gamma=\{x_2=y_2=0\}$ is written in the following form
$$V=Q_1(x_1,x_2)x_2^3+\lambda_1x_2^2+\lambda_0,\quad \lambda_0,\lambda_1\in \mathbf{C}.$$
\end{crl}

\subsubsection{NVE reducible to quantum harmonic oscillator}

  We have seen, that hamiltonians \eqref{:Hqarmonic}, has
generic NVE along curves in $\Gamma$ of type (\ref{:quadratic}).
Once again, we apply theorem \ref{polynint}.

\begin{prp}
Hamiltonians of the family \eqref{:Hqarmonic} do not admit any
additional rational first integral.
\end{prp}

  We can also discuss, the Picard-Vessiot integrability of those
equations \eqref{:quadratic}. First, we shall notice that by just
an scaling of $t$,
$$t = \frac{\tau}{\sqrt[4]{c_2}}-\frac{c_1}{2c_2}$$
we reduce it to an \emph{quantum harmonic oscillator equation},

\begin{equation}\label{:E}
\frac{d^2\xi}{d\tau^2} = (\tau^2 - E)\xi, \quad\quad E =
\frac{c_1^2-4c_0c_1}{4\sqrt{c_2^3}}.
\end{equation}

  In appendix (remark \ref{:rkm2}) we analyze this equation. It is
\emph{Picard-Vessiot integrable} if and only if $E$ is an odd positive number. Then, let
us compute de parameter $E$ associated to NVE of integral curves of
Hamiltonians \ref{:Hqarmonic}.

  Let us keep in mind that the family of those curves is
parameterized by
  $$h = \frac{y_1^2}{2}+\frac{\lambda_4}{(\lambda_2+2\lambda_3x_1)^2}$$
  In order to fix the parameterization of those curves, let us assume
that time $t=0$ corresponds to $x_1 = 0$. The NVE corresponding to a
curve, depending on energy $h$, is written:
  $$\ddot \xi = (c_0(h)+c_1(h)t + c_2(h)t^2)\xi.$$
We compute these coefficients $c_i(h)$ using,
$$y_1 = \frac{\sqrt{2h(\lambda_2+2\lambda_3x_1)^2-2\lambda_4}}{\lambda_2+2\lambda_3x_1}
\xrightarrow{t\to
0}\frac{\sqrt{2h\lambda_2^2-2\lambda_4}}{\lambda_2},$$ and then, by
applying the hamiltonian field,
$$c_0(h) = \frac{\lambda_1}{2},\quad c_1(h) =
\sqrt{\frac{h\lambda_2^2-\lambda_4}{2}},\quad c_2(h) = \lambda_3h,$$
and then, it is reducible to equation (\ref{:E}) with parameter,
$$E = \frac{1}{8\sqrt{\lambda_3^3}}\left(\frac{\lambda_2^2-4\lambda_1\lambda_3}{\sqrt{h}}-\frac{\lambda_4}{\sqrt{h^3}} \right),$$

If $\lambda_2 = 4\lambda_1\lambda_3$ and $\lambda_4 = 0$, then parameter $E$ vanish for every
integral curve in $\Gamma$.  For any other case, $E$ is a non-constant analytical function of $h$.

\medskip

\emph{We have proven that those  NVE are, generically not Picard-Vessiot integrable for any hamiltonian
of the \eqref{:Hqarmonic} family.}

\subsubsection{NVE Mathieu}

  In order to apply theorem \ref{:MR}, we just need to make some remarks on the field
of coefficients. Let $\gamma$ be a generic integral curve of
\eqref{:Mhamiltonian1}, or \eqref{:Mhamiltonian2}. Those curves
are, in general, Riemann spheres. The field of coefficients
$\mathcal M_\gamma$, is generated by $x_1, y_1$, so that it is
$\mathbf C(x_1,\dot x_1)$. We also have,
  $$a = \lambda_0 + \lambda_1 x_1 + \lambda_2 x_1^2,  \quad \dot a = \dot x_1(\lambda_1 + 2\lambda_2 x_1),$$
so that, for $\lambda_2 = 0$, we have
  $$\mathcal M_\gamma = \mathbf(\alpha,\dot\alpha) = \mathbf C(\sin t, cos t) = \mathbf \C(e^{it}).$$
and, for $\lambda_2 \neq 0$,
  $$\mathbf \C(e^{it}) \hookrightarrow \mathcal M_\gamma $$
is an algebraic extension.

\medskip

  So that, in our algebrization algorithm, the field of coefficients of Mathieu equation is taken,
$\mathbf \C(e^{it})$. For $\lambda_2 = 0$, we can apply directly
theorem \ref{:MR}, and for $\lambda_2 \neq 0$, we can apply theorem
\ref{moramis} and then theorem \ref{:MR}.

\medskip

  Non trivial equations of type Mathieu, with field of coefficients
$\mathbf C(e^{it})$, analyzed in remark \ref{Mathiewobs}, have Galois
group $SL(2,\mathbf C)$. Thus for computed families of hamiltonians
with NVE of type Mathieu, we get:

\begin{prp}
  Hamiltonians of the families \eqref{:Mhamiltonian1} and \eqref{:Mhamiltonian2}
if $\lambda_1 \neq 0$ and $(\lambda_1,\lambda_2) \neq (0,0)$
respectively, do not admit any additional rational first integral.
\end{prp}

\subsection*{Thanks and acknowledgements}

 We want to thank Juan Morales-Ruiz for
his valuable help, advices, suggestions, and also for proposing the
initial problem. We also acknowledge Sergi Sim\'on for his valuable
suggestions in the first stage of developing our method. We are also
indebted with Jackes Arthur Weil by his suggestions on Kovacic's
algorithm.

 This research is supported by project {\bf BFM2003-09504-C02-02}
(Spanish Government).


\section*{Appendix}

\appendix

\section{Kovacic's Algorithm}
This algorithm is devoted to solve the reduced linear differential
equation (RLDE) $\xi''=r\xi$ and is based on the algebraic
subgroups of $SL(2,\mathbf{C}).$ For more details see \cite{Kov}.
Improvements for this algorithm is given in \cite{UlmerWeil}, in
where is not necessary to reduce the equation. Here, we follows
the original version given by Kovacic.

\begin{thm}\label{subgroups} Let $G$ be an algebraic subgroup of $SL(2,\mathbf{C})$.  Then
one of the following four cases can occur.
\begin{enumerate}
\item $G$ is triangularizable.
\item $G$ is conjugate to a subgroup of infinite dihedral group (also called meta-abelian group)
and case 1 does not hold.
\item Up to conjugation $G$ is either of following finite groups: Tetrahedral group, Octahedral
group or Icosahedral group, and cases 1 and 2 do not hold.
\item $G = SL(2,\mathbf{C})$.
\end{enumerate}
\end{thm}

Each case in the Kovacic algorithm is related with each one of the
algebraic subgroups of $SL(2,\mathbf{C})$ and the associated
Riccatti equation
$$\theta^{\prime}=r-\theta ^{2}=\left( \sqrt{r}-\theta\right)
\left(  \sqrt{r}+\theta\right),\quad\theta={\xi'\over \xi}.$$

According to theorem \ref{subgroups}, there are four cases in the
Kovacic algorithm. Only for cases 1, 2 and 3 we can solve the
differential equation RLDE, but for the case 4 we have not
Liouvillian solutions for RLDE. Is possible that the Kovacic
algorithm only can provide us only one solution ($y_1$), so that
we can obtain the second solution ($y_2$) through
\begin{equation}\label{second}
y_2=y_1\int\frac{dx}{y_1^2}.
\end{equation}

{\bf\large Notations.} For the equation RLDE with
$$r={s\over t},\quad s,t\in \mathbf{C}[x]$$
we use the following notations.
\begin{enumerate}
\item Denote by $\Gamma'$ be
the
set of (finite) poles of $r$, $\Gamma^{\prime}=\left\{  c\in\mathbf{C}%
:t(c)=0\right\}$.

\item Denote by
$\Gamma=\Gamma^{\prime}\cup\{\infty\}$.
\item By the order of $r$ at
$c\in \Gamma'$, $\circ(r)_c$, we mean the multiplicity of $c$.

\item By the order of $r$ at $\infty$, $\circ\left(
r_{\infty}\right) ,$ we mean the order of $\infty$ as a zero of
$r$. That is $\circ\left( r_{\infty }\right) =deg(t)-deg(s)$.

\item By the order of $r$ at
$c\in \Gamma'$, $\circ(r)_c$, we mean the multiplicity of $c$.

\end{enumerate}
\subsection{The four cases}

{\bf\large Case 1.} In this case $\left[ \sqrt{r}\right] _{c}$ and
$\left[ \sqrt{r}\right] _{\infty}$ means the Laurent series of
$\sqrt{r}$ at $c$ and the Laurent series of $\sqrt{r}$ at $\infty$
respectively. Furthermore, we define $\varepsilon(p)$ as follows:
if $p\in\Gamma,$ then $\varepsilon\left( p\right) \in\{+,-\}.$
Finally, the complex numbers $\alpha_{c}^{+},\alpha_{c}^{-},\alpha_{\infty}%
^{+},\alpha_{\infty}^{-}$ will be defined in the first step. If
the differential equation has not poles it only can fall in this
case.
\medskip

{\bf Step 1.} Search for each $c \in \Gamma'$ and for $\infty$ the
corresponding situation such as follows:

\medskip

\begin{description}

\item[$(c_{0})$] If $\circ\left(  r_{c}\right)  =0$, then
$$\left[ \sqrt {r}\right] _{c}=0,\quad\alpha_{c}^{\pm}=0.$$

\item[$(c_{1})$] If $\circ\left(  r_{c}\right)  =1$, then
$$\left[ \sqrt {r}\right] _{c}=0,\quad\alpha_{c}^{\pm}=1.$$

\item[$(c_{2})$] If $\circ\left(  r_{c}\right)  =2,$ and $$r= \cdots
+ b(x-c)^{-2}+\cdots,\quad \text{then}$$
$$\left[ \sqrt {r}\right]_{c}=0,\quad \alpha_{c}^{\pm}=\frac{1\pm\sqrt{1+4b}}{2}.$$

\item[$(c_{3})$] If $\circ\left(  r_{c}\right)  =2v\geq4$, and $$r=
(a\left( x-c\right)  ^{-v}+...+d\left( x-c\right)
^{-2})^{2}+b(x-c)^{-(v+1)}+\cdots,\quad \text{then}$$ $$\left[
\sqrt {r}\right] _{c}=a\left( x-c\right) ^{-v}+...+d\left(
x-c\right) ^{-2},\quad\alpha_{c}^{\pm}=\frac{1}{2}\left(
\pm\frac{b}{a}+v\right).$$

\item[$(\infty_{1})$] If $\circ\left(  r_{\infty}\right)  >2$, then
$$\left[\sqrt{r}\right]  _{\infty}=0,\quad\alpha_{\infty}^{+}=0,\quad\alpha_{\infty}^{-}=1$$

\item[$(\infty_{2})$] If $\circ\left(  r_{\infty}\right)  =2,$ and
$r= \cdots + bx^{2}+\cdots$, then $$\left[
\sqrt{r}\right]  _{\infty}=0,\quad\alpha_{\infty}^{\pm}=\frac{1\pm\sqrt{1+4b}%
}{2}$$

\item[$(\infty_{3})$] If $\circ\left(  r_{\infty}\right) =-2v\leq0$,
and
$$r=\left( ax^{v}+...+d\right)  ^{2}+ bx^{v-1}+\cdots,\quad \text{then}$$
$$\left[  \sqrt{r}\right]  _{\infty}=ax^{v}+...+d,\quad
\text{and}\quad \alpha_{\infty}^{\pm }=\frac{1}{2}\left(
\pm\frac{b}{a}-v\right).$$
\end{description}
\medskip

{\bf Step 2.} Find $D\neq\emptyset$ defined by
$$D=\left\{
m\in\mathbf{Z}_{+}:m=\alpha_{\infty}^{\varepsilon
(\infty)}-%
{\displaystyle\sum\limits_{c\in\Gamma^{\prime}}}
\alpha_{c}^{\varepsilon(c)},\forall\left(  \varepsilon\left(
p\right) \right)  _{p\in\Gamma}\right\}  .$$ If $D=\emptyset$,
then we should start with the case 2. Now, if $\#D>0$, then for
each $m\in D$ we search $\omega$ $\in\mathbf{C}(x)$ such that
$$\omega=\varepsilon\left(
\infty\right)  \left[  \sqrt{r}\right]  _{\infty}+%
{\displaystyle\sum\limits_{c\in\Gamma^{\prime}}}
\left(  \varepsilon\left(  c\right)  \left[  \sqrt{r}\right]  _{c}%
+{\alpha_{c}^{\varepsilon(c)}}{(x-c)^{-1}}\right).$$
\medskip

{\bf Step 3}. For each $m\in D$, search for a monic polynomial
$P_m$ of degree $m$ with
$$P_m'' + 2\omega P_m' + (\omega' + \omega^2 - r) P_m = 0.$$

If success is achieved then $\xi_1=P_m e^{\int\omega}$ is a
solution of the differential equation RLDE.  If not, then case 1
cannot hold.
\bigskip

{\bf\large Case 2.}  Search for each $c \in \Gamma'$ and for
$\infty$ the corresponding situation such as follows:
\medskip

{\bf Step 1.} Search for each $c\in\Gamma^{\prime}$ and $\infty$
the sets $E_{c}\neq\emptyset$ and $E_{\infty}\neq\emptyset.$ For
each $c\in\Gamma^{\prime}$ and for $\infty$ it is define
$E_{c}\subset\mathbf{Z}$ and $E_{\infty}\subset\mathbf{Z}$ as
follows:
\medskip

\begin{description}
\item[($c_1$)] If $\circ\left(  r_{c}\right)=1$, then $E_{c}=\{4\}$

\item[($c_2$)] If $\circ\left(  r_{c}\right)  =2,$ and $r= \cdots +
b(x-c)^{-2}+\cdots ,\ $ then $$E_{c}=\left\{
2+k\sqrt{1+4b}:k=0,\pm2\right\}.$$

\item[($c_3$)] If $\circ\left(  r_{c}\right)  =v>2$, then $E_{c}=\{v\}$

\item[$(\infty_{1})$] If $\circ\left(  r_{\infty}\right)  >2$, then
$E_{\infty }=\{0,2,4\}$

\item[$(\infty_{2})$] If $\circ\left(  r_{\infty}\right)  =2,$ and
$r= \cdots + bx^{2}+\cdots$, then $$E_{\infty }=\left\{
2+k\sqrt{1+4b}:k=0,\pm2\right\}.$$

\item[$(\infty_{3})$] If $\circ\left(  r_{\infty}\right)  =v<2$,
then $E_{\infty }=\{v\}$
\medskip
\end{description}

{\bf Step 2.} Find $D\neq\emptyset$ defined by
$$D=\left\{
m\in\mathbf{Z}_{+}:\quad m=\frac{1}{2}\left(  e_{\infty}-
{\displaystyle\sum\limits_{c\in\Gamma^{\prime}}} e_{c}\right)
,\forall e_{p}\in E_{p},\text{ }p\in\Gamma\right\}.$$ If
$D=\emptyset,$ then we should start the case 3. Now, if $\#D>0,$
then for each $m\in D$ we search a rational function $\theta$
defined by
$$\theta=\frac{1}{2}
{\displaystyle\sum\limits_{c\in\Gamma^{\prime}}}
\frac{e_{c}}{x-c}.$$
\medskip

{\bf Step 3.} For each $m\in D,$ search a monic polynomial $P_m$
of degree $m$, such that
$$P_m^{\prime\prime\prime}+3\theta
P_m^{\prime\prime}+(3\theta^{\prime}+3\theta
^{2}-4r)P_m^{\prime}+\left(  \theta^{\prime\prime}+3\theta\theta^{\prime}%
+\theta^{3}-4r\theta-2r^{\prime}\right)P_m=0.$$ If $P_m$ there is
not exists, then the case 2 cannot hold. If such a polynomial is
found, set $\phi = \theta + P'/P$ and let $\omega$ be a solution
of
$$\omega^2 + \phi \omega + {1\over2}\left(\phi' + \phi^2 -2r\right)=
0.$$

Then $\xi_1 = e^{\int\omega}$ is a solution of the differential
equation RLDE.
\bigskip

{\bf\large Case 3.} Search for each $c \in \Gamma'$ and for
$\infty$ the corresponding situation such as follows:
\medskip

{\bf Step 1.} Search for each $c\in\Gamma^{\prime}$ and $\infty$
the sets $E_{c}\neq\emptyset$ and $E_{\infty}\neq\emptyset.$ For
each $c\in\Gamma^{\prime}$ and for $\infty$ it is define
$E_{c}\subset\mathbf{Z}$ and $E_{\infty}\subset\mathbf{Z}$ as
follows:
\medskip

\begin{description}

\item[$(c_{1})$] If $\circ\left(  r_{c}\right)  =1$, then
$E_{c}=\{12\}$

\item[$(c_{2})$] If $\circ\left(  r_{c}\right)  =2,$ and $r= \cdots +
b(x-c)^{-2}+\cdots$, then
\begin{displaymath}
E_{c}=\left\{ 6+k\sqrt{1+4b}:\quad
k=0,\pm1,\pm2,\pm3,\pm4,\pm5,\pm6\right\}.
\end{displaymath}

\item[$(\infty)$] If $\circ\left(  r_{\infty}\right)  =v\geq2,$ and $r=
\cdots + bx^{2}+\cdots$, then
$$E_{\infty }=\left\{
6+{12k\over n}\sqrt{1+4b}:\quad
k=0,\pm1,\pm2,\pm3,\pm4,\pm5,\pm6\right\},\quad n\in\{4,6,12\}.$$
\medskip
\end{description}

{\bf Step 2.} Find $D\neq\emptyset$ defined by
$$D=\left\{
m\in\mathbf{Z}_{+}:\quad m=\frac{n}{12}\left(
e_{\infty}-{\displaystyle\sum\limits_{c\in\Gamma^{\prime}}}
e_{c}\right)  ,\forall e_{p}\in E_{p},\text{
}p\in\Gamma\right\}.$$ In this case we start with $n=4$ to obtain
the solution, afterwards $n=6$ and finally $n=12$. If
$D=\emptyset$, then the differential equation has not Liouvillian
solution because falls in the case 4. Now, if $\#D>0,$ then for
each $m\in D$ with its respective $n$, it is search a rational
function
$$\theta={n\over 12}{\displaystyle\sum\limits_{c\in\Gamma^{\prime}}}
\frac{e_{c}}{x-c}$$ and a polynomial $S$ defined as $$S=
{\displaystyle\prod\limits_{c\in\Gamma^{\prime}}} (x-c).$$

{\bf Step 3}. Search for each $m\in D$, with its respective $n$, a
monic polynomial $P_m=P$ of degree $m,$ such that its coefficients
can be determined recursively by
$$\bigskip P_{-1}=0,\quad P_{n}=-P,$$
$$P_{i-1}=-SP_{i}^{\prime}-\left( \left( n-i\right)
S^{\prime}-S\theta\right)  P_{i}-\left( n-i\right)  \left(
i+1\right)  S^{2}rP_{i+1},$$ where $i\in\{0,1\ldots,n-1,n\}.$ If
$P$ there is not exists, then the differential equation has not
Liouvillian solution because falls in the case 4. Now, if $P$
there exists, it is search $\omega$ such that $$
{\displaystyle\sum\limits_{i=0}^{n}} \frac{S^{i}P}{\left(
n-i\right)  !}\omega^{i}=0,$$ then a solution of the differential
equation RLDE is given by $$\xi=e^{\int \omega},$$ where $\omega$
is solution of the previous polynomial of degree $n$.
\bigskip

\subsection{Some remarks on Kovacic's algorithm}
Along of this section we assume that RLDE falls only in one of the
four cases.
\begin{obs}[Case 1]\label{rkov2}
If RLDE fall in case 1, then its Galois group is given by any of
the following groups:

\begin{description}

\item[I1] $e$ when the algorithm provides two
rational solutions or only one rational solution and the second
solution obtained by equation \eqref{second} has not logarithmic
term.
$$e=\left\{\begin{pmatrix}1&0\\0&1\end{pmatrix}\right\},$$
this group is connected and abelian.

\item[I2] $\mathbf{G}_k$ when the algorithm provides only one
algebraic solution $\xi$ such that $\xi^k\in\mathbf{C}(x)$ and
$\xi^{k-1}\notin\mathbf{C}(x)$.
$$\mathbf{G}_k=\left\{\begin{pmatrix}\lambda&d\\0&\lambda^{-1}\end{pmatrix}:\quad \lambda\text{  is a $k$-root of the unity, }d\in\mathbf{C}\right\},$$
this group is disconnected and its identity component is abelian.

\item[I3] $\mathbf{C}^*$ when the algorithm provides two
non algebraic solutions.
$$\mathbf{C}^*=\left\{\begin{pmatrix}c&0\\0&c^{-1}\end{pmatrix}:c\in \mathbf{C}^*\right\},$$
this group is connected and abelian.

\item[I4] $\mathbf{C}^{+}$ when the algorithm provides one
rational solution and the second solution is not algebraic.
$$\mathbf{C}^{+}=\left\{\begin{pmatrix}1&d\\0&1\end{pmatrix}:d\in \mathbf{C}\right\}, \quad {\xi}\in\mathbf{C}(x),$$
this group is connected and abelian.

\item[I5] $\mathbf{C}^*\ltimes\mathbf{C}^{+}$ when the algorithm only provides one
solution $\xi$ such that $\xi$ and its square are not rational
functions.
$$\mathbf{C}^*\ltimes\mathbf{C}^{+}=\left\{\begin{pmatrix}c&d\\0&c^{-1}\end{pmatrix}:c\in\mathbf{C}^*,d\in \mathbf{C}\right\}, \quad \xi\notin\mathbf{C}(x),\quad {\xi}^2\notin\mathbf{C}(x).$$
This group is connected and non-abelian.

\item[I6] $SL(2,\mathbf{C})$ if the algorithm do not provides any
solution. This group is connected and non-abelian.
\end{description}
\end{obs}

\begin{obs}[Case 2]If RLDE fall in case 2, then the Kovacic's Algorithm can provides one or two
solutions. This depends of $r$ such as follows:
\begin{description}
\item[II1] if $r$ is given by
$$r={2\phi'+2\phi-\phi^2\over 4},$$
then there exists only one solution,

\item[II2] if $r$ is given by
$$r\neq{2\phi'+2\phi-\phi^2\over 4},$$
then there exists two solutions.
\item[II3] The identity component of the Galois group for this
case is abelian.
\end{description}
\end{obs}

\begin{obs}[Case 3]If RLDE falls in case 3, then its Galois group is given by any of
the following groups:
\begin{description}
\item[III1] {\bf Tetrahedral group} when $\omega$ is obtained with $n=4.$ This group of order 24 is generated by
$$
\begin{pmatrix}
e^{\frac{k\pi i}{3}} & 0\\
0 & e^{-\frac{k\pi i}{3}}
\end{pmatrix}
, \quad\frac{1}{3}\left(  2e^{\frac{k\pi i}{3}}-1\right)
\begin{pmatrix}
1 & 1\\
2 & -1
\end{pmatrix},\quad k\in \mathbf{Z}.$$
\item[III2] {\bf Octahedral group} when $\omega$ is obtained with $n=6.$ This group of order 48 is generated by
$$
\begin{pmatrix}
e^{\frac{k\pi i}{4}} & 0\\
0 & e^{-\frac{k\pi i}{4}}
\end{pmatrix}, \quad \frac{1}{2}e^{\frac{k\pi i}{4}}\left(  e^{\frac{k\pi
i}{2}}+1\right)
\begin{pmatrix}
1 & 1\\
1 & -1
\end{pmatrix},\quad k\in \mathbf{Z}.$$

\item[III3]{\bf Icosahedral group} when $\omega$ is obtained with $n=12.$ This group of order 120 is generated by
$$
\begin{pmatrix}
e^{\frac{k\pi i}{5}} & 0\\
0 & e^{-\frac{k\pi i}{5}}
\end{pmatrix}
,\quad
\begin{pmatrix}
\phi & \psi\\
\psi & -\phi
\end{pmatrix},\quad k\in \mathbf{Z},$$ being $\phi$ and $\psi$ defined as
$$\phi=\frac{1}{5}\left(  e^{\frac{3k\pi
i}{5}}-e^{\frac{2k\pi i}{5} }+4e^{\frac{k\pi i}{5}}-2\right),
\quad \psi=\frac{1}{5}\left( e^{\frac{3k\pi i}{5}}+3e^{\frac{2k\pi
i}{5}}-2e^{\frac{k\pi i}{5}}+1\right)$$

\item[III4] The identity component of the Galois group for this
case is abelian.
\end{description}
\end{obs}

\normalsize

\vspace{0.8cm}


\vspace{0.6cm}\small

\rightline{\sc Departament de Matem\`atica Aplicada II}
\rightline{\sc Universitat Polit\`ecnica de Catalunya}
\rightline{\sc C. Jordi Girona, 1-3}
\rightline{\sc Barcelona,
Espanya}
\smallskip
\rightline{{\it e-mail:} \tt primitivo.acosta@upc.edu}
\smallskip
\rightline{{\it e-mail:} \tt david.blazquez@upc.edu}

\end{document}